# Monte Carlo Study of a Spin-3/2 Blume-Emery-Griffiths Model on a Honeycomb Lattice

M. Žukovič*, M. Jaščur

Institute of Physics, Faculty of Sciences, P. J. Šafárik University, Park Angelinum 9, 041 54 Košice, Slovakia

We study the phase diagram of the spin-3/2 Blume-Emery-Griffiths model on a honeycomb lattice by Monte Carlo simulations in order to verify the presence of some peculiar features predicted by the effective field theory (EFT) with correlations. The locations of the order-disorder phase boundaries are estimated from thermal variations of the magnetic susceptibility curves. It is found that for positive values of the biquadratic interactions the critical boundary shows a discontinuous character as a function of the single-ion anisotropy strength, in line with the EFT expectations. However, for negative values of the biquadratic interactions the step-like variation of the critical frontier predicted by EFT was not reproduced.

### 1. Introduction

The spin-3/2 Blume-Emery-Griffiths (BEG) model is a spin-3/2 Ising model which includes both bilinear and biquadratic interactions, as well as single-ion uniaxial crystal-field anisotropy. It was introduced to understand behavior of some real physical systems, such as ternary mixtures. The model has been intensively studied by various approaches (e.g., [1-6]), nevertheless, its critical behavior is far from being well understood. Even in the most studied case with zero biquadratic interactions, i.e., the Blume-Capel (BC) model, no consensus among various approaches has been established regarding whether the first-order line separating (1/2,1/2) and (3/2,3/2) ferromagnetic phases at low temperatures extends to the disordered phase boundary line or it terminates at an isolated point [3,4].

The spin-3/2 BEG model with finite biquadratic interactions was much less investigated [3,5,6]. The calculations by an effective field theory with correlations (EFT) for the system on a honeycomb lattice [5] produced some peculiar features that could not be observed in the spin-3/2 BC model. Particularly interesting was the step-wise dependence of the phase boundary on the single-ion anisotropy parameter for larger negative values of the biquadratic interactions strength.

In the present investigations we employ Monte Carlo (MC) simulations with the aim to verify whether the features predicted by the EFT, in particular the step-wise behavior of the phase boundary, are real or just artifacts of the used approximation.

### 2. Model and methods

We consider the spin-3/2 BEG model on a honeycomb lattice described by the Hamiltonian

$$H = -J_1 \sum_{\langle i,j \rangle} S_i S_j - J_2 \sum_{\langle i,j \rangle} S_i^2 S_j^2 - D \sum_i S_i^2, \quad (1)$$

where $S_i = \pm 1/2, \pm 3/2$ is a spin on the i-th lattice site, $\langle i,j \rangle$ denotes the sum over nearest neighbors, $J_1 > 0$ is a ferromagnetic bilinear exchange interaction parameter, $J_2$ is a biquadratic exchange interaction parameter and $D$ is a single-ion anisotropy parameter.

The honeycomb lattice system is considered to consist of two interpenetrating sublattices A and B. Then, assuming sublattice uniformity we can focus on an elementary unit cell comprising the central spin, let say from the sublattice A, i.e., $S_A$, and its three nearest neighbors from the sublattice B, i.e., $S_B$, and express its reduced ground-state (GS) energy per spin as

$$e/J_1 = -(3S_A S_B + 3\alpha S_A^2 S_B^2 + \Delta(S_A^2 + S_B^2))/2, \quad (2)$$

where $\alpha = J_2/J_1$ and $\Delta = D/J_1$. Then, in the parameter space $(\alpha,\Delta)$, GS is one of the states $(\pm 1/2, \pm 1/2)$, $(\pm 1/2, \pm 3/2)$, $(\pm 3/2, \pm 3/2)$, that minimizes expression (2).

In order to study thermal behavior of the model and to determine its phase diagram, we perform MC simulations on a spin system of a moderate linear size of $L = 48$, employing the Metropolis dynamics and applying the periodic boundary conditions. For thermal averaging we consider $N = 5 \times 10^4$ MCS (Monte Carlo sweeps) after discarding another $10^4$ MCS for thermalization. For configurational averaging and estimating the error bars we carry out three independent runs. The simulations are performed within a wide range of the values of the reduced single-ion anisotropy parameter $\Delta$ by varying the reduced temperature $k_B T/J_1$. For each value of $\Delta$ the simulations start from high temperatures in the paramagnetic region, using a random initial configuration, and then the temperature is gradually decreased with the step $\Delta k_B T/J_1 = 0.05$ and the simulation starts from the final configuration obtained at the previous temperature value. We calculate the total magnetization per site

$$m = \langle M \rangle / L^2 = \langle \sum_{i=1}^{L^2} S_i \rangle / L^2 \quad (3)$$

and the corresponding magnetic susceptibility

*corresponding author; e-mail: milan.zukovic@upjs.sk

$$\chi = \frac{\langle M^2 \rangle - \langle M \rangle^2}{L^2 k_B T}. \quad (4)$$

The total magnetization serves as an order parameter, i.e., it takes finite values in the region of long-range magnetic ordering and vanishes in the paramagnetic region. The transition points can be located from the susceptibility maxima.

## 3. Results and discussion

The temperature dependencies of the magnetization curves for different values of the single-ion anisotropy parameter $\Delta$ are presented in Fig. 1(a), for a selected value of $\alpha = -2$. The plots demonstrate that the low-temperature region is long-range ordered for any value of $\Delta$ and becomes disordered at higher temperatures. Note that due to finiteness of the lattice $m$ remains finite even in the paramagnetic region and becomes zero only in the thermodynamic limit of $L \to \infty$. However, the character of the low-temperature ordered phase depends on $\Delta$. As discussed above, taking the value of $\alpha = -2$, the system is in GS ($\pm 1/2, \pm 1/2$) if $\Delta < 3/4$, in GS ($\pm 1/2, \pm 3/2$) if $3/4 < \Delta < 45/4$, and in GS ($\pm 3/2, \pm 3/2$) if $\Delta > 45/4$. The susceptibility curves corresponding to the magnetization dependences in Fig. 1(a) are plotted in Fig. 1(b). The sharp peaks appear at the phase transition points.

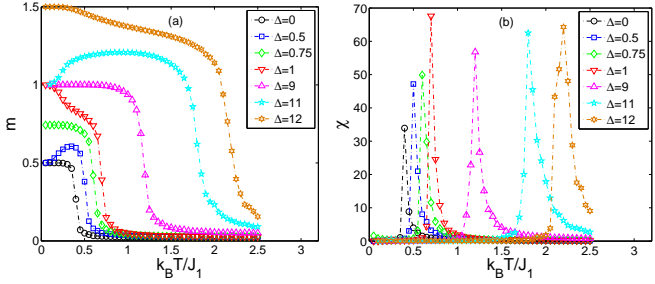

Fig.1. Temperature variations of (a) the magnetization and (b) the susceptibility, for various values of the single-ion anisotropy parameter $\Delta$ and $\alpha = -2$.

The resulting order-disorder phase diagram, determined from the susceptibility peaks locations, is shown in Fig. 2, for different values of the exchange interaction ratio $\alpha$. The error bars represent standard deviations of the peaks locations obtained from multiple MC runs. In the phase diagrams we can observe that for positive values of $\alpha$, such as $\alpha = 2$, the MC results confirm the discontinuous character of the critical line as a function of the single-ion anisotropy strength observed in the EFT study [5]. However, for negative values of $\alpha \leq -1$, the non-monotonic step-like variation of the critical frontier was not reproduced. Namely, only two steps associated with the phase transitions from the state ($\pm 1/2, \pm 1/2$) to ($\pm 1/2, \pm 3/2$) and then to ($\pm 3/2, \pm 3/2$) were observed.

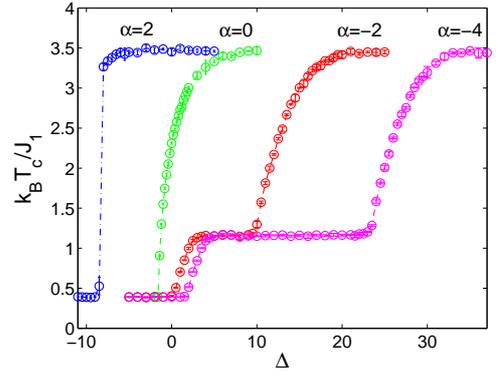

Fig.2. Phase diagrams for selected values of $\alpha$. The areas below and above the curves represent magnetically ordered and disordered phases, respectively.

## 4. Conclusions

We determined the phase diagram of the spin-3/2 BEG model on a honeycomb lattice by MC simulations with the goal to understand some unexpected features, such as the step-wise behavior of the phase boundary, obtained in the earlier study by EFT. For positive values of the biquadratic interactions our results confirmed the discontinuous character of the critical line as a function of the single-ion anisotropy strength. However, for larger negative values of the biquadratic interactions the step-like variation of the critical frontier was not confirmed. Therefore, the multiple steps in the EFT results are considered artifacts of the used approximation.

### Acknowledgement

This work was supported by the Scientific Grant Agency of Ministry of Education of Slovak Republic (Grant No. 1/0234/12). The authors acknowledge the financial support by the ERDF EU (European Union European Regional Development Fund) grant provided under the contract No. ITMS26220120047 (activity 3.2.).